\documentclass[preprint, amsmath,amssymb, aps]{revtex4-2}
\usepackage{graphicx}
\usepackage{dcolumn}
\usepackage{bm}
\usepackage{xcolor}
\usepackage{lineno}
\usepackage{ulem}
\newcommand{\avg}[1]{\left\langle #1  \right\rangle}

\begin{document}
\preprint{APS/123-QED}

\title{Stochastic model for football's collective dynamics}

\author{A. Chacoma}
\email{achacoma@famaf.unc.edu.ar}
\affiliation{Instituto de F\'isica Enrique Gaviola (IFEG-CONICET)}
\affiliation{Facultad de Matem\'atica, Astronom\'ia, F\'isica y Computaci\'on, Universidad Nacional de C\'ordoba}

\author{N. Almeira}
\affiliation{Instituto de F\'isica Enrique Gaviola (IFEG-CONICET)}
\affiliation{Facultad de Matem\'atica, Astronom\'ia, F\'isica y Computaci\'on, Universidad Nacional de C\'ordoba}

\author{J.I. Perotti}
\affiliation{Instituto de F\'isica Enrique Gaviola (IFEG-CONICET)}
\affiliation{Facultad de Matem\'atica, Astronom\'ia, F\'isica y Computaci\'on, Universidad Nacional de C\'ordoba}

\author{O.V. Billoni}
\affiliation{Instituto de F\'isica Enrique Gaviola (IFEG-CONICET)}
\affiliation{Facultad de Matem\'atica, Astronom\'ia, F\'isica y Computaci\'on, Universidad Nacional de C\'ordoba}

\begin{abstract}
In this paper, we study collective interaction dynamics emerging in the game of football--soccer.
To do so, we surveyed a database containing body--sensors traces measured during three professional football matches, where we observed statistical patterns that we used to propose a stochastic model for the players' motion in the field.
The model, which is based on linear interactions, captures in good approximation the spatiotemporal dynamics of a football team. 
Our theoretical framework, therefore, can be an effective analytical tool to uncover the underlying cooperative mechanisms behind the complexity of football plays.
Moreover, we showed that it can provide handy theoretical support for coaches to evaluate teams' and players' performances in both training sessions and competitive scenarios.

\end{abstract}

\maketitle

\section{Introduction}

The use of complex systems theory as an alternative paradigm for analyzing elite sport dynamics is currently arousing intense academic interest \cite{almeira2017structure,schaigorodsky2014memory,perotti2013innovation, erkol2020best}.
Fostered by the new advances in data acquisition \cite{pappalardo2019public,pettersen2014soccer} and artificial intelligence techniques \cite{fister2015computational,neiman2011reinforcement,mukherjee2019prior}, the use of state--of--the--art statistical tools to evaluate teams' performances is, nowadays, shaping a new profile of data--driven--based professional coaches worldwide.

One can find in the literature plentiful research works in several fields of physics that have been devoted to studying phenomena related to sports science \cite{laksari2018mechanistic,schade2012influence,le2015theory,yu2016textured,trenchard2014collective}.
The research community in statistical physics, focused in the study of sports mainly in the framework of the stochastic processes.
For instance, studying the time evolution of scores \cite{clauset2015safe,ribeiro2012anomalous,kiley2016game}.
Alternatively, other studies propose innovative models, based on ordinary differential equation \cite{ruth2020dodge}, stochastic agent-based game simulations \cite{chacoma2020modeling}, and network science theory \cite{yamamoto2021preferential}, aiming to describe the complex dynamical behavior of teams' players.

Formally, sports teams can be thought of as complex {\it sociotechnical systems} \cite{davids2013complex}, where a wide range of organizational factors might interact to influence athletes' performances \cite{hulme2019sports,petersen2020renormalizing,mclean2017s,soltanzadeh2016systems}.
Particularly, in collective games like football, cooperative interplay dynamics seem to be a key feature to be analyzed \cite{gudmundsson2017review, gonccalves2016effects}.
In principle, collective behaviors in soccer are important since they are connected to team tactics and strategies. 
Usually, features of these collective behaviors are described by using simple group-level metrics~\cite{clemente2013collective, frencken2011oscillations,
moura2013spectral, yue2008mathematicalp1, yue2008mathematicalp2,narizuka2019clustering}.
Furthermore, temporal sequences of ball and player movements in football, showing traits of complex
behaviors, have been reported and studied using stochastic models and statistical analysis~\cite{mendes2007statistics, kijima2014emergence, narizuka2017chasing,chacoma2020modeling}.

Recent works has focused on describing cooperative on--ball interaction in football within the framework of network science \cite{martinez2020spatial,herrera2020pitch,buldu2019defining,gonccalves2017exploring,narizuka2021networks}. 
In \cite{garrido2020consistency}, for instance, D. Garrido et al. studied the so call {\it Pitch Passing Networks} in the games of the Spanish League at $2018/2019$ season. 
In this outstanding work, the authors use network metrics and topological aspects to define teams' consistency and identifiability, two highly relevant global indicators to analyze team performance during competition.

From an alternative perspective, our research group has focused on studying the dynamical interactions at the microscopic level, i.e. by modeling player--player interactions. 
In a previous work \cite{chacoma2020modeling}, we proposed an agent--based model that correctly reproduces key features of global statistics indicators of nearly $2000$ real--life football matches.
In the same line, the present paper aims to describe the complexity of this game uncovering the underlying mechanisms ruling the collective dynamics.
Our goal is to give a new step towards a full description of the spatiotemporal dynamics in a football match.
With this purpose, we posit a simple model based on linear interactions to describe teams' cooperative evolution.
To do so, we analyzed a public database containing body--sensor traces from three professional football matches of the Norwegian team {\it Tromsø IL} (see section \ref{se:metodos}). 
We will show that our model succeeds in capturing part of the cooperative dynamics among the players and that higher--order contributions (non--linear interactions) can be carefully modeled as fluctuations.
Moreover, we will show that our framework provides a useful tool to analyze and evaluate tactical aspects of the teams. 

This paper is divided into three sections: in section Material and Methods, we describe the database, the statistical regularities found in the data, and formally propose our theoretical framework. 
In section Results and Discussion, we present our main results and relevant findings. Our conclusion and future perspectives are briefly summarized in the last section.

\section{Material and methods}
\label{se:metodos}
\subsection{The database}

In the year 2014, S.A. Pettersen et al. published a database of traces recorded with body--sensors from three professional soccer games recorded in November 2013 at the Alfheim Stadium in Troms{\o}, Norway \cite{pettersen2014soccer}.
This database is divided into five datasets, each one contains the halves of the game Troms{\o}~IL vs. Str{\o}msgodset~IF ($DS1$, $DS2$), the halves of the game Troms{\o}~IL vs. Anzhi Makhachkala ($DS3$, $DS4$), and $40$ minutes of the game Troms{\o}~IL vs. Tottenham Hotspurs ($DS5$).
This contribution also offers video records, but they were not used in our research. 
We highlight that only Troms{\o}~IL gave consent to be tracked with the body--sensors, thus the traces available in the datasets are only from this team. The goalkeeper's position, likewise, was not tracked. 

The player positions were measured at $20~Hz$ using the highly accurate {\it ZXY Sport Tracking system} by {\it ChyronHego} (Trondheim, Norway). 
However, to perform our analysis, we pre--processed the data so as to have the players' position in the field in one-second windows.
In this way, we lose resolution but it becomes simpler to analyze players' simultaneous movements or coordination maneuvers.

\subsection{Statistical regularities}

\begin{figure*}[t!]
\centering
\includegraphics[width=1\textwidth]{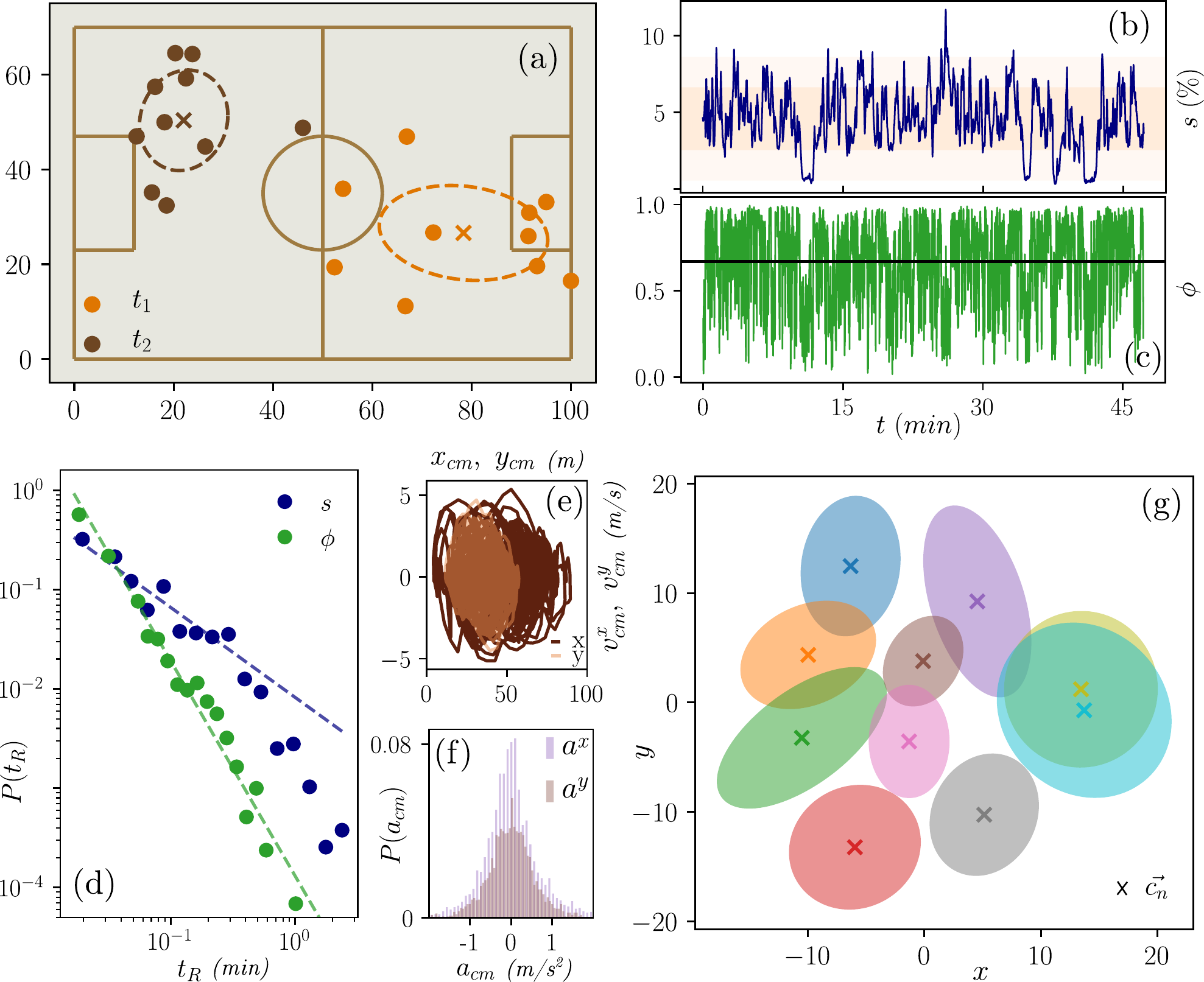}
\caption{
Statistical regularities in $DS1$.
(a) The team deployed in the field at $t_1=7~(min)$ and $t_2=8~(min)$. Ellipses are a measure of the characteristic area of the team at both times (see main text).
(b) Evolution of the characteristic area $s$. 
The width of the narrow bands around the curve equals the values of $1\times \sigma_s$ and $2\times \sigma_s$, where $\sigma_s$ is the standard deviation.
(c) Evolution of the order parameter $\phi$. Black solid horizontal line indicates the mean value of the series at $\approx 0.7$.
(d) Time to return to the mean value calculated from series $s$ and $\phi$. Dashed lines indicate a non--linear fit performed to measure the power--law exponent (see main text).
(e) The dynamics of the center of mass.
Position vs. velocity on both axis.
(f) Distribution of instantaneous accelerations of the center of mass in both axis.
(g) Action zones for the players (ellipses) in the center of mass frame of reference. 
$\vec{c_n}$ indicates each player center (see main text for further details).
}
\label{fi:stat}
\end{figure*}

In this section, we describe relevant aspects of the statistical observations that we used to propose our stochastic model. In this case, we focus on the analysis of $DS1$, but similar results can be obtained for the others datasets.
Let us discuss Fig.~\ref{fi:stat}. In panel $(a)$ we show the team deployed in the field at two different times. 
In an offensive position at $t_1= 7~(min)$, and in an defensive position at $t_2=8~(min)$. 
The ellipses, drawn in the figure with dashed lines, give the standard deviation intervals around the mean (see supplementary material S1 for further details) and can be thought of as a measure of the team's dispersion in the field.
The ellipses' area can be thought of as a characteristic area of the team deployed in the field.
Note, the system seems to suffer an expansion at $t_1$ and a contraction at $t_2$.
To characterize this process we study the temporal evolution of 
$s:= 100\times\frac{Ellipse~Area}{Field~area}$ ($\%$), that we show in panel $(b)$. 
We measured $\avg{s(t)} = 5.1$, $\sigma_s = 2.2$. 
The low dispersion in the time series and the symmetry around the mean, indicate the system moves around the field with a well--defined characteristic area exhibiting small variations throughout the dynamics of the game.

Let us now focus on analyzing the level of global ordering in the team.
To do so, we analyze the evolution of the parameter $\phi (t) = \big | \frac{1}{N}\sum^N_{n=1} \frac{\vec{v}_n(t)}{|\vec{v}_n(t)|} \big|$, (see eq.~(1) in \cite{cavagna2010scale}).
Where $v_n$ is the velocity of player $n$, and $N$ is the total number of players.
In the case $\phi \approx 1$, this parameter indicates the players move as a {\it flock}, following the same direction.
On the contrary, when $\phi \approx 0$, they move in different directions.
%
%
Panel $(c)$, shows the temporal evolution of $\phi$.
We measured $\avg{\phi(t)}= 0.7$, $\sigma_{\phi} = 0.2$. 
This level of global ordering during the game shows that there are time intervals when the players tend to move as a highly coordinated {\it flock} \cite{welch2021collective}. 
Therefore, it seems there are interactions among the teammates that cause the emergence of global ordering.

We now turn our attention to the analysis of the temporal structure of series $s(t)$ and $\phi(t)$.
Let us define $t_R$ as the time of return to the mean value.
In panel $(d)$ we show the distribution $P_{s}(t_R)$ and $P_{\phi}(t_R)$.
In both cases, we can see heavy--tailed distributions. 
By performing a non--linear fit, using the expression $P(t_R) = C t_R^{-\gamma}$, for the case of the relative area, s, within the range $(0, 0.4)~min$,
we measured $\gamma_s = 0.96 \pm 0.03$.
In the case of the order parameter $\phi$, in the whole range, we obtained $\gamma_{\phi} = 2.16 \pm 0.04$.
It is well known that for a random walk process in one dimension the probability of first return turns out $\gamma= 3/2$ \cite{redner2001guide}. 
In our case, the measured non--trivial exponents seem to indicate the presence of a complex multiscale temporal structure in the dynamics.
Notice that these two parameters, related to the team structure and order, wrap somehow the memory and complex dynamics for the team during the match.

We now focus on describing the dynamics of the center of mass (CM).
In Fig.~\ref{fi:stat} panel $(e)$, we show the relations $x_{cm}$ vs. $v^x_{cm}$, and $y_{cm}$ vs. $v^y_{cm}$. 
We can see the position variables are bounded to the field area, and the velocities in both axes seem to be bounded within the small range  $(-5,5)$ ($m/s$).
In panel $(f)$, on the other hand, we show the distribution of accelerations. We measured $\avg{a^x_{cm}}=\avg{a^y_{cm}}=0$, $\sigma_{a^x_{cm}}=0.4$, and $\sigma_{a^y_{cm}}=0.2$ ($m/s^2$).
Since the CM of the system is barely accelerated, we can approximate the center of mass as an inertial system. Then, to simplify our analysis, we can study the dynamical motion of the players from the center of mass frame of reference.
In this frame, we aim to define action zones for the players.
To do so, we analyze the positions in the plane that the players have explored during the match. 
The ellipses in panel $(g)$, give the standard--deviation intervals around the mean, and can be thought of as characteristic action zones for the players.
Note that, from this perspective, it naturally emerges that Troms{\o}~IL used in this half the classical tactical system {\it 4-4-2}.

Summarizing, we observed $(i)$ the spatial dispersion of the players follows a well defined characteristic area, $(ii)$ inside the team, the players' movements exhibit correlation and global ordering, $(iii)$ the system shows a complex multiscale temporal structure, and $(iv)$ the players' motion can be studied from the center of mass frame of reference, simplifying the analysis.
In the next section, we use these insights to propose a simple stochastic model of cooperative interaction to analyze the spatiotemporal evolution of the team during the match.

\subsection{The model}

The interplay among teammates can be thought of as individuals that cooperate to run a tactical system.
In this frame, we aim to define a model to describe the spatiotemporal evolution of the team.
Since our goal is to define a simple theoretical framework such that we can easily interpret the results, we propose a model based on players to players' interactions. 
We proceed as follows (i) we define the equations of evolution for the players in the team, (ii) we use the empirical data to fit the equations' parameters, and (iii) we model the error in the fitting as fluctuation in the dynamics.
In the following, we present the results in this regard.

\subsubsection{The equations of evolution}

In the CM frame of reference, we define 
$\vec{r}_n(t)= \big(x_n(t), y_n(t)\big)^T$ and 
$\vec{v}_n(t)= \big(v_n^x(t), v_n^y(t)\big)^T$ 
as the position and velocity of player $n$ at time $t$.
We propose the dynamical variables change driven by interactions that can be thought of as springs--like forces.
Every player in our model is bounded to $(i)$ a place in the field related to their natural position in the team, $\vec{a}_n$, and $(ii)$ the other players.

The equation of motion for a team player $n$ can be written as follow,
\begin{equation}
\centering
M_n \ddot{\vec{r}}_n = 
-\gamma_n\vec{v}_n 
+ k_{an}(\vec{a}_n-\vec{r}_n) +  
{\sum_m}^\prime k_{nm}(\vec{r}_m-\vec{r}_n),
\label{eq:eqmov}
\end{equation}
where the first term is a damping force, the second one is an ``anchor'' to the player's position, and the sum is the contribution of the interaction forces with the other players. 
We propose different interaction constants in the horizontal and vertical axis, therefore the parameters $\gamma_n$, $k_{an}$, and $k_{nm}$, are $2-D$ diagonal matrices such as, 
$\gamma_n= 
\big(\begin{smallmatrix}
  \gamma_n^x & 0\\ 
  0 & \gamma_n^y
\end{smallmatrix}\big)$,
$k_{an}= 
\big(\begin{smallmatrix}
  k_{an}^x & 0\\ 
  0 & k_{an}^y
\end{smallmatrix}\big)$,
$k_{nm}= 
\big(\begin{smallmatrix}
  k_{nm}^x & 0\\ 
  0 & k_{nm}^y
\end{smallmatrix}\big)$.
Moreover, since players have comparable weights, for simplicity we consider $M_n=1$ for all the players.
Notice, if equilibria exist, $\vec{r}_n(t\rightarrow \infty)= \vec{r}_n^*$, and $\vec{v}_n(t\rightarrow \infty)=0$.
Then,
\begin{equation}
-\big(k_{na}+ {\sum_m}^\prime k_{nm}\big) {\vec{r^*}_n}
+ {\sum_m}^\prime k_{nm}\vec{r^*}_m
+ k_{an}\vec{a}_n=0,
\label{eq:equilibrium}
\end{equation}
must hold. 

Eqs. \ref{eq:eqmov} can also be written as a first order equations system as follows,
\begin{equation}
\begin{split}
\dot{\vec{r}}_n &= \vec{v}_n\\
\dot{\vec{v}}_n &= 
-\big(k_{an}+{\sum_m}^\prime k_{nm}\big) \vec{r}_n
+ {\sum_m}^\prime k_{nm}\vec{r}_m
-\gamma_n \vec{v}_n 
+ k_{an}\vec{a}_n.
\end{split}
\label{eq:system1}
\end{equation}
Furthermore, by defining $\vec{x}=\big(x_1,..x_{n}, v^x_1,...,v^x_{n}\big)$ and $\vec{y}=\big(y_1,..y_{n}, v^y_1,...,v^y_{n}\big)$, we can write,

\begin{equation}
\begin{split}
\dot{\vec{x}} = J^x \big(\vec{x}-\vec{x^*}\big)\\
\dot{\vec{y}} = J^y 
\big(\vec{y}-\vec{y^*}\big),
\end{split}
\label{eq:jacobian}
\end{equation}
where $J^x, ~ J^y \in R^{2n \times 2n}$ are the Jacobian matrices of system (\ref{eq:system1}). 
With Eqs.~(\ref{eq:jacobian}), we can analyze separately the system evolution along the horizontal and the vertical axis. Moreover, in the section \ref{se:results} we will show that the Jacobian matrices can be used to describe the team's collective behavior.

\subsubsection{Fitting the model's parameters}
\label{se:fitting}

In this section we show how to fit the model's parameters $\gamma_n$, $k_{an}$, $k_{nm}$, and $\vec{a}_n$ by using the datasets, and Eqs.~(\ref{eq:system1}) and (\ref{eq:equilibrium}).
To proceed, we have considered the following steps,
\begin{enumerate}
    \item For every player in the team, each dataset provides the position in the field $\vec{r}_n(t)$. The velocity is calculated as $\vec{v}_n(t):=\frac{\vec{r}_n(t+\Delta t)- \vec{r}_n(t)}{\Delta t}$ ($\Delta t=1~s$)
    
    \item The discrete version of system (\ref{eq:system1}), gives the tool to estimate the states of the players at time $t+\Delta t$ by using as inputs the real states at time $t$ and the model's parameters,
    \begin{equation*}
    \begin{split}
    \vec{r}_n(t+\Delta t)^{\prime} &=
    \vec{r}_n(t)+\vec{v}_n(t)\Delta t\\
    \vec{v}_n(t+\Delta t)^{\prime} &= \vec{v}_n(t)+ \\
    &+ \big(-\gamma_n \vec{v}_n(t)-
    \big(k_{an}+{\sum_m}^\prime k_{nm}\big) \vec{r}_n(t)+ 
    \\
    &+{\sum_m}^\prime k_{nm}\vec{r}_m(t)+ k_{an}\vec{a}_n\big) \Delta t.
    \end{split}
    \label{eq:discretesystem}
    \end{equation*}
    Where $\vec{r}_n(t+\Delta t)^{\prime}$ and $\vec{v}_n(t+\Delta t)^{\prime}$ are the model's estimations.
    
    \item Note, by considering the definition of the velocity made in $1$, $\vec{r}_n(t+\Delta t) = \vec{r}_n(t+\Delta t)^{\prime}$. 
    Then, at every step, the parameters are only used to predict the new velocities.
\end{enumerate}
Moreover, to simplify our framework, 
\begin{enumerate}
    \setcounter{enumi}{3}
    \item 
    Since parameters $\vec{a}_n$ are linked to the equilibria values by Eq.~(\ref{eq:equilibrium}), 
    without lost of generality, we take $\vec{c}_n=\vec{r^*}_n$ (where $\vec{c_n}$ is the center of the action radii for every players, empirically obtained from the datasets see Fig. \ref{fi:stat}~D). 
    By doing this, values $\vec{a}_n$ can be calculated from the values of $\vec{c}_n$ and the other parameters.
    
    \item 
    To simplify our analysis, we normalized the players' position in the dataset such that the standard deviation (scale) of players' velocities are the unit (i.e. $\sigma_{\vec{v}}=1$). This is useful for later assess the fitting performance.
\end{enumerate}

In this frame, we define the error $\vec{\xi}_n(t) := \vec{v}_n(t+\Delta t)-\vec{v}_n(t+\Delta t)^\prime $, and fit $\gamma_n$,  $k_{an}$, $k_{nm}$ by minimizing the sum $\sum_t \sum_n \big|\vec{\xi}_n(t)\big|$.
With this method, we obtain a unique set of parameters that govern the equations.
For a more detailed description of the minimizing procedure, c.f. supplementary material S2.

Notice that, to avoid possible large fluctuations linked to drastic tactical changes, we fitted the set of parameters to each dataset (a half match). 
This criterion, at the same time, let us compare the strength of the interactions among different matches halves. 

The results of the optimization process for $DS1$ are given in Table~\ref{tabla1}. 
There we show the values of the fitted parameters in both coordinates, $x$, and $y$.
We can see $\alpha_n \approx 10^{-1}$, $k_{an} \approx 10^{-2}$ and $k_{nm} \lesssim 10^{-2}$.
Particularly interesting are parameters $k_{nm}$, since they indicate the strength of the interactions among players.
In this case, we can see a wide variety of values, from strong interactions as in the case of players $1-2$ to negligible interactions in the case of players $3-8$.
For results on other datasets, please c.f. supplementary material S3.

\subsubsection{Modeling $\vec{\xi}_n(t)$ as fluctuations in the velocities}

By using the optimal set of parameters calculated with the method proposed in the previous section, we can calculate for all the players at every time step the difference between the real velocity and the model's prediction. This defines $N$ temporal series $\vec{\xi}_n(t) = \big \{\vec{\xi}_n(t_0), \vec{\xi}_n(t_1), ...,\vec{\xi}_n(t_T)\big\}$, that can be thought of as stochastic fluctuations in the players velocities.
Note that, in the context of a football match, these fluctuations can be related to stochastic forces acting upon the players. 
With this idea in mind, we propose to introduce in the system (\ref{eq:system1}) a noisy component linked to these fluctuations.
With this aim, we focus on analyzing and describing the behavior of $\vec{\xi}_n(t)$.

\begin{figure*}[t!]
\centering
\includegraphics[width=1\textwidth]{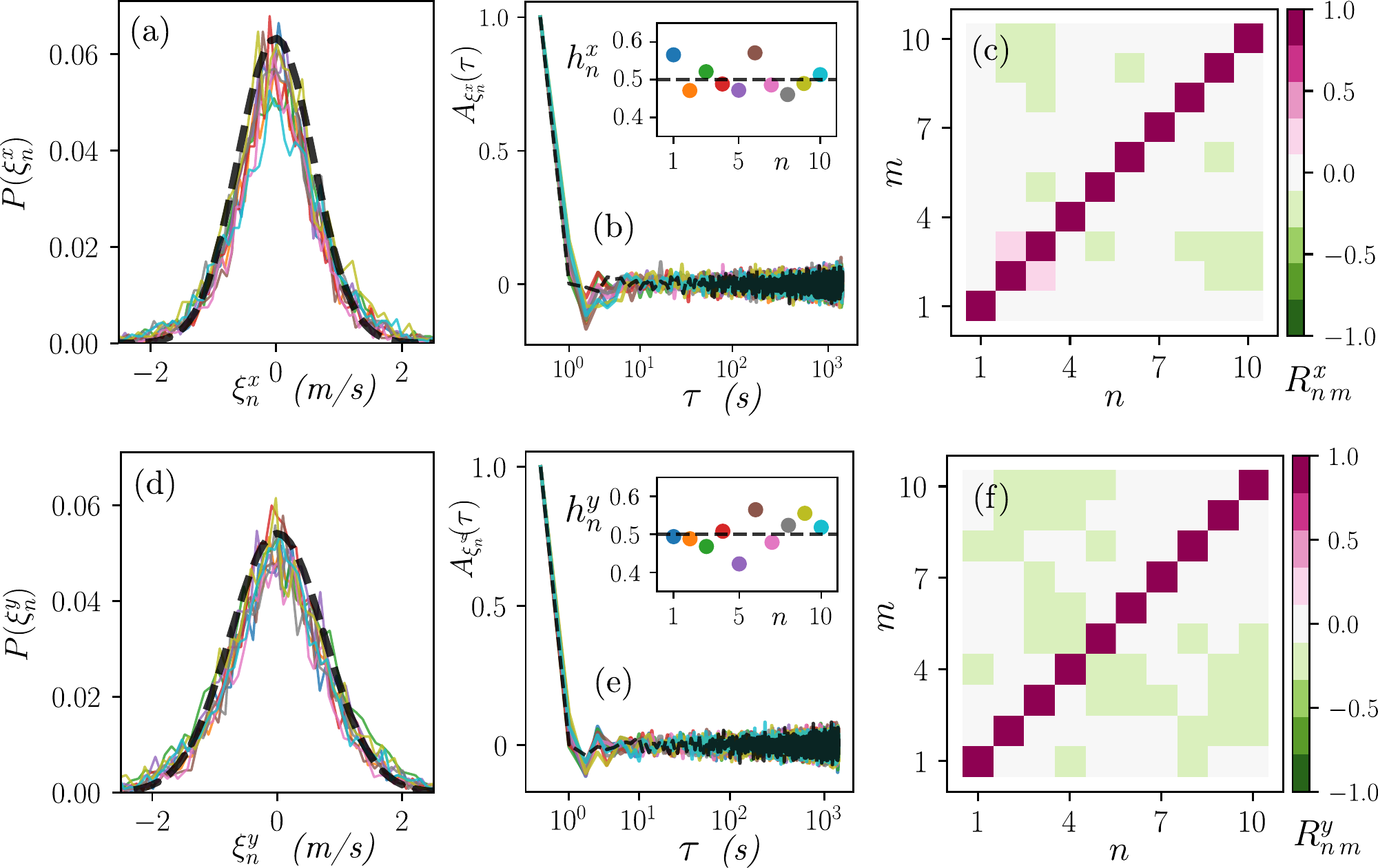}
\caption{
Error characterization -- Fluctuation analysis.
For the horizontal axis, $x$.
(a) Distributions of the error values for the ten players (colored curves). Dashed lines indicate a Gaussian fit performed to the aggregated set of values.
(b) Main: autocorrelation functions for the ten cases (colored curves). The dashed line indicates the same calculation for a white noise process. Inset: Hurst exponent for the ten cases. 
(c) Pearson matrix indicating the value of the linear correlation among each pair of time series, including auto--correlations.
For the vertical axis, $y$, the descriptions of panels (d), (e) and (f) are analogous to the previous case.
}
\label{fi:ruido}
\end{figure*}

Let us turn our attention to Fig.~\ref{fi:ruido}. 
Here, our goal is to characterize the fluctuations linked to $DS1$.
In panel $(a)$, we show the distributions of values related to $\xi^x_n$ ($n=1,..,10$). We can see, in each case, the curves approach a Gaussian shape.
Dashed line indicate a non--linear fit performed to the distribution given by the entire set of values ($\xi^x$), in this case, we have measured $\avg{\xi^x} = 0.001 \pm 0.002$ and $\sigma_{\xi^x} =0.60 \pm 0.02$ $(m/s)$ ($R^2=0.97$).
Note, the fluctuations scale is lower than the velocities scale $\sigma_{\xi^x}<\sigma_v$ (with $\sigma_v=1$, c.f. section \ref{se:fitting} item $5$). 
Panel $(b)$, on the other hand, shows the autocorrelation functions 
$A_{\xi_n^x}(\tau) = 
\frac{\avg{Z Z^\prime} - \avg{Z} \avg{Z^\prime}}
{\sigma_{Z} \sigma_{Z^{\prime}}}$, 
with $Z=\xi^x_n(t)$ and $Z^\prime =\xi^x_n(t+\tau)$.
For each case, we can see an abrupt decay at the beginning. 
To help the eye to visualize the behavior of the curves, the black dashed line in the plot shows the autocorrelation function for a white noise process.
The inset in the panel shows the values of the Hurst exponent calculated by performing a Detrended Fluctuation Analysis (DFA) to $\xi^x_n(t)$.
We obtain values around $0.5 \pm 0.06$, which is consistent with a set of memoryless processes.
Panel $(c)$, on the other hand, shows the Pearson matrix $R_{nm}^x = \frac{C_{nm}^x}{\sqrt{C_{nn}^xC_{mm}^x}}$, where $C_{nm}^x$ is the covariance matrix of series $\xi^x_n(t)$.
We can see that $R_{nm}^x< 0.25~\forall~n,m$; which indicates a low level of linear correlation among the fluctuations associated with the different players.
A similar description with analogous results can be done for $\xi^y_n$, by analyzing panels $(d)$, $(e)$ and $(f)$.

Base on the observations made above, we propose to model the fluctuations in both axis as non-correlated Gaussian noise, such that 
$\vec{\xi}_n(t)= \vec{\sigma}_n \xi_n$, with
$\avg{\xi_n(t)}=0$,
$\avg{\xi_n(t)\xi_n(t^\prime)}= \delta (t-t^\prime)$, 
$\avg{\xi_n(t)\xi_m(t)}= 0$, 
and $\vec{\sigma_n}= (\sigma^x_n, \sigma^y_n)$  the empirical measured scales for the processes.


\section{Results and discussion}
\label{se:results}

\subsection{Simulations on the collective dynamics}

As we previously stated, we firstly used the datasets to fit the model's parameters, and secondly characterized the errors as fluctuation in the velocities.
With these inputs, in the frame of our model, we can simulate the players' collective dynamics and compare the results with empirical data to assess the model performance.
In order to do this, we modify system (\ref{eq:system1}) as follow, 

\begin{equation}
\begin{split}
d\vec{r}_n &= \vec{v}_n dt\\
d\vec{v}_n &= \big[
-\big(k_{na}+{\sum_m}^\prime k_{nm}\big) \vec{r}_n
+ {\sum_m}^\prime k_{nm}\vec{r}_m -
\\
&-\gamma_n \vec{v}_n 
+ k_{na}\vec{a}_n 
\big] dt + \vec{dW_n},
\end{split}
\label{eq:system2}
\end{equation}
where $d\vec{W}_n = \vec{\sigma}_n \xi_n dt$, with $\vec{\sigma}_n$ and $\xi_n dt$ as were defined in the previous section.
Note, (\ref{eq:system2}) is a system of stochastic differential equation (SDE). To solve it, we use the Euler--Maruyama algorithm for Ito equations.
In this section, we show the results linked to the dataset $DS1$, similar results for the other datasets can be found in the supplementary material S4.

Let us focus on Fig.~\ref{fi:simulations}.
Panels $(a)$ and $(b)$ show two heatmaps with the probability of finding a team player in the position $(x,y)$.
The left panel shows the results for the empirical data, whereas the right panel for simulations.
For better visualization, in both cases, the probabilities were normalized to the maximum value, defining the parameter $\rho \in [0,1]$.
As we can observe, simulations reproduce reasonably well the empirical observations. 
To quantify the result, we calculated the Jensen--Shannon distance ($DJS$) between the distributions. We measured $DJS= 0.05$, which indicates a good similarity between distributions.
In panel $(c)$, on the other hand, we compare players' action zones. The empirical observations are the shadow ellipses, whereas the simulation the curves. 
We can see, on the whole, the model gives a good approximation. Particularly, for those areas with smaller dispersion.
In panel $(d)$, we analyze the kinetic energy of the system,
$E_k := \sum_ n \frac{1}{2} |\vec{v}_n|^2$.
Our goal here is to globally describe the temporal structure of the system.
In the inset, we show the temporal evolution of this quantity. Regarding the mean values of the energy we measured $\avg{E_k}_{DS1}=10.3$ for the data, and $\avg{E_k}_{MO}=11.5$ for simulations. 
We can see, the energy reaches high peaks in the empirical case that are not observed in the outcomes of the model compensating for the slightly lower mean. 
This effect indicates the presence of higher--order contributions, probably linked to non--linear interactions, that our model based on linear interactions does not take into account.
The main plot in the panel shows the distribution of the times to return to the mean value for the kinetic energy, $P(t_R)$, for both empirical data and model.
We can see in both cases, $P(t_R) \propto t_R^{-\gamma_{E_k}}$,  with $\gamma_{E_k} = 2.4 \pm 0.1$.
Note, the value of the exponent, agree we the value measured for the time to return in the case of the variable $\phi$ (see section Materials and Methods, Statistical regularities). 
This seems to indicate that the temporal structure captured by the evolution of the energy, may be related to the emergence of order in the system.
Let us turn our attention to the distributions' tails. For the case of the empirical data, we can see the presence of extreme events that are not observed in simulations. This effect, as well as the peaks in $E_k$, could be linked to higher--order contributions of the interactions. 

To summarize, in this section, we showed that despite its simplicity, our first--order stochastic model succeeds in uncovering several aspects of the complexity of the spatiotemporal structure underlying the dynamics.
In the next two sections, we show how to take profit of the fitted interactions to describe the players' individual and collective behavior.

\subsection{Describing the team behavior by analyzing the model's parameters}

\subsubsection{Hierarchical clustering classification on the team lineup}

\begin{figure*}[t!]
\centering
\includegraphics[width=\textwidth]{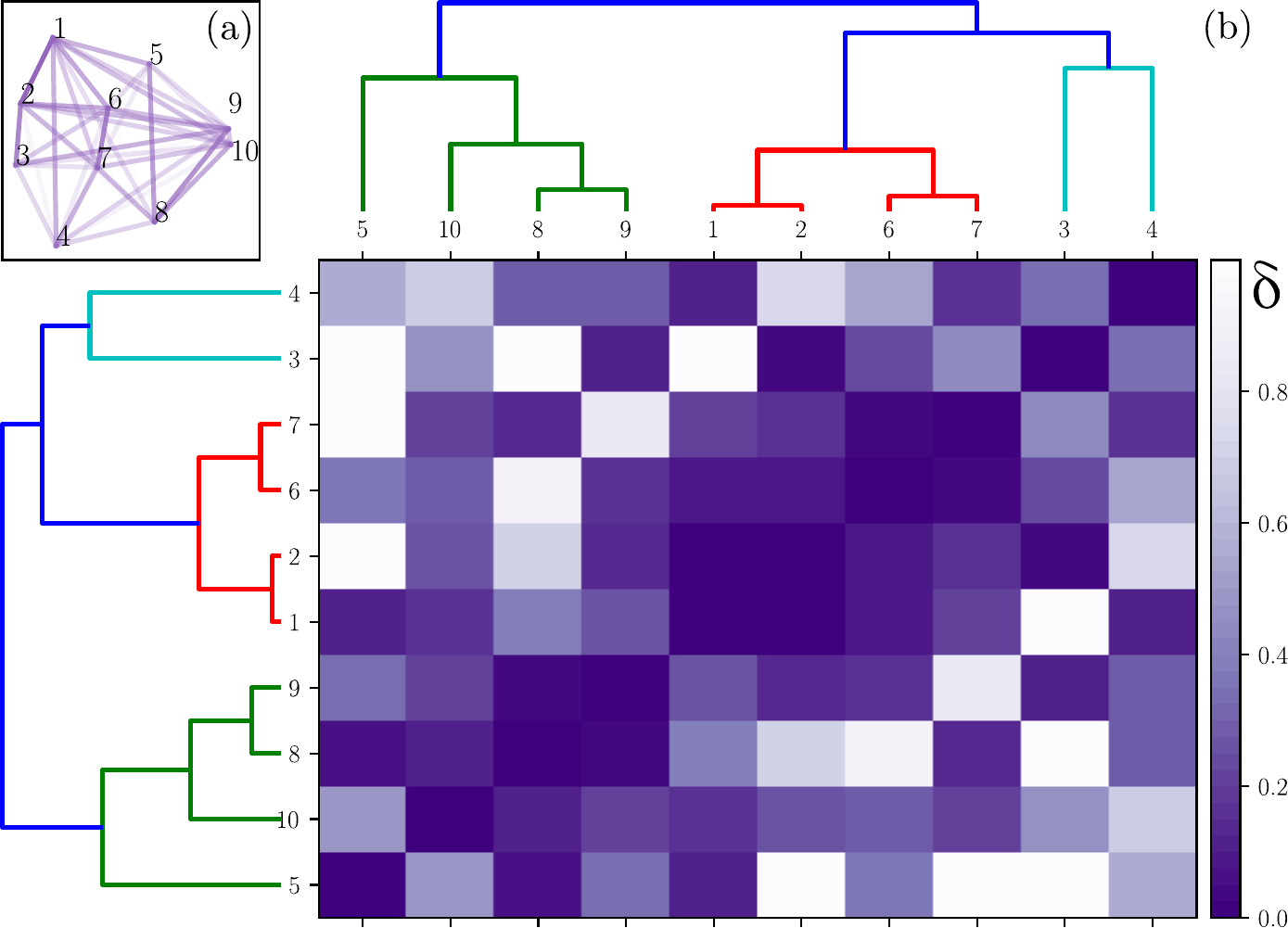}
\caption{
Classification of the team members, based on the inferred interaction strengths.
(a) Visualization of the connection strength, $k_{nm}$, among the players.
(b) Hierarchical relations among the players. At the center: the distance matrix. At the top and the left: Dendrograms to visualize the clusters of players within the team.
}
\label{fi:conexiones}
\end{figure*}

The interaction parameters $\vec{k}_{nm}$ can be useful to analyze the interplay among team players, and to describe the environment they are constrained to. 
In Fig.~\ref{fi:conexiones} panel $(a)$, we show a visualization of the players' interactions magnitude.
In this networks of players, the links' transparency (alpha value), represents the connections strength between players $n$ and $m$. These values are calculated as $\kappa_{nm} = \sqrt{({k^x}_{nm})^2 + ({k^y}_{nm})^2}$.
Note, the connection values can be used as a proxy of the distance among the players. 
Let us define the distance 
$\delta_{nm}:= e^{-\frac{\kappa_{nm}}{\sigma_\kappa}}$, between player $n$ and $m$, where $\sigma_\kappa$ is the standard deviation of the set of values $\kappa_{nm}$.
Note, the exponential function in $\delta_{nm}$ is used to define large distances linked to small connections, and short distances related to strong connections.
Then, based on this metric, we calculate the matrix of distances. 
With this matrix, by using a hierarchical clustering classification technique, we detect small communities of players within the team.
In the following, we discuss the results in this regard.
In Fig.~\ref{fi:conexiones} panel $(b)$, we show the re--ordered matrix of distances with two equal dendrograms, one placed at the top and the other at the left, showing the hierarchical relationship between players.
To perform this calculation we used the {\it ward} method \cite{ward1963hierarchical}.
With this classification, we can easily observe the presence of two main clusters of players.
Those colored in green, players $5-10-8-9$, can be related to the offensive part of the team; the others to the defensive.
Within the latter group, the cluster colored in red (players $1-2-6-7$) are related to back and middle defenders at the left side of the field, whereas players $3-4$ to back defenders at the right.
Within the former group, we can differentiate between a central--right group of attackers, $8,9,10$, and an individual group giving by player $5$. 
As we said above, this technique allows us to study groups of players with strong interactions during the match. 
For instance, let us focus on analyzing the group of players $6-7$. 
These players cover the central zone, they are likely in charge of covering the gaps when other defenders go to attack. For instance, the advance of the wing--backs $1-4$ by the sides.
We should expect them to behave similarly, which agrees with the result of our classification.  
The case of player $5$ is particularly interesting.
Our results indicate that this player is less constrained than the others attackers. 
Probably, our classification frame is detecting that $5$ is a free player in the team, a classic {\it play builder} on the midfield, in charge of generating goal--scoring opportunities. 

The information provided by the hierarchical clustering classification, allows us to characterize the players' behavior within a team and, therefore, provide useful insight into the collective organization.
In the light of this technique, it is possible to link the strengths and weaknesses of the team to the levels of coordination among the players. 
For instance, if it is observed a lack in the levels of coordination among the rival players at the right side of their formation (as we can see in the case of Troms{\o}~IL in $DS1$), it would be interesting to foster attacks in this sector.

We could also perform a comparative study by analyzing several games to correlate results with levels of coordination among the players. 
With this approach, we can detect if there are patterns related to a winning or a losing formation, providing valuable information for coaches to use.
The same idea can be applied to training sessions, to promote routines oriented to strengthen the players' connection within particular groups in the line-up, aiming to improve the team performance in competitive scenarios.

To summarize, the hierarchical clustering analysis evidences the presence of highly coordinated behavior among subgroups of players, that can be directly related to their role in the team. 
In this framework, coaches may find a useful tool to support the complex decisions--making process involved in the analysis of the tactical aspects of the team, assess players' performances, propose changes, etc.

\subsubsection{Collective modes}

The eigenvalues and eigenvectors of the Jacobian matrices $J^x$ and $J^y$ (see Eqs.~(\ref{eq:jacobian})) can be handy to describe some aspects of the team players dynamics on the field.
Note, if the system exhibit complex eigenvalues the eigenvectors bears information on the collective modes of the system, and, consequently, on the collective behavior of the team.

\begin{figure*}[t!]
\centering
\includegraphics[width=0.5\textwidth]{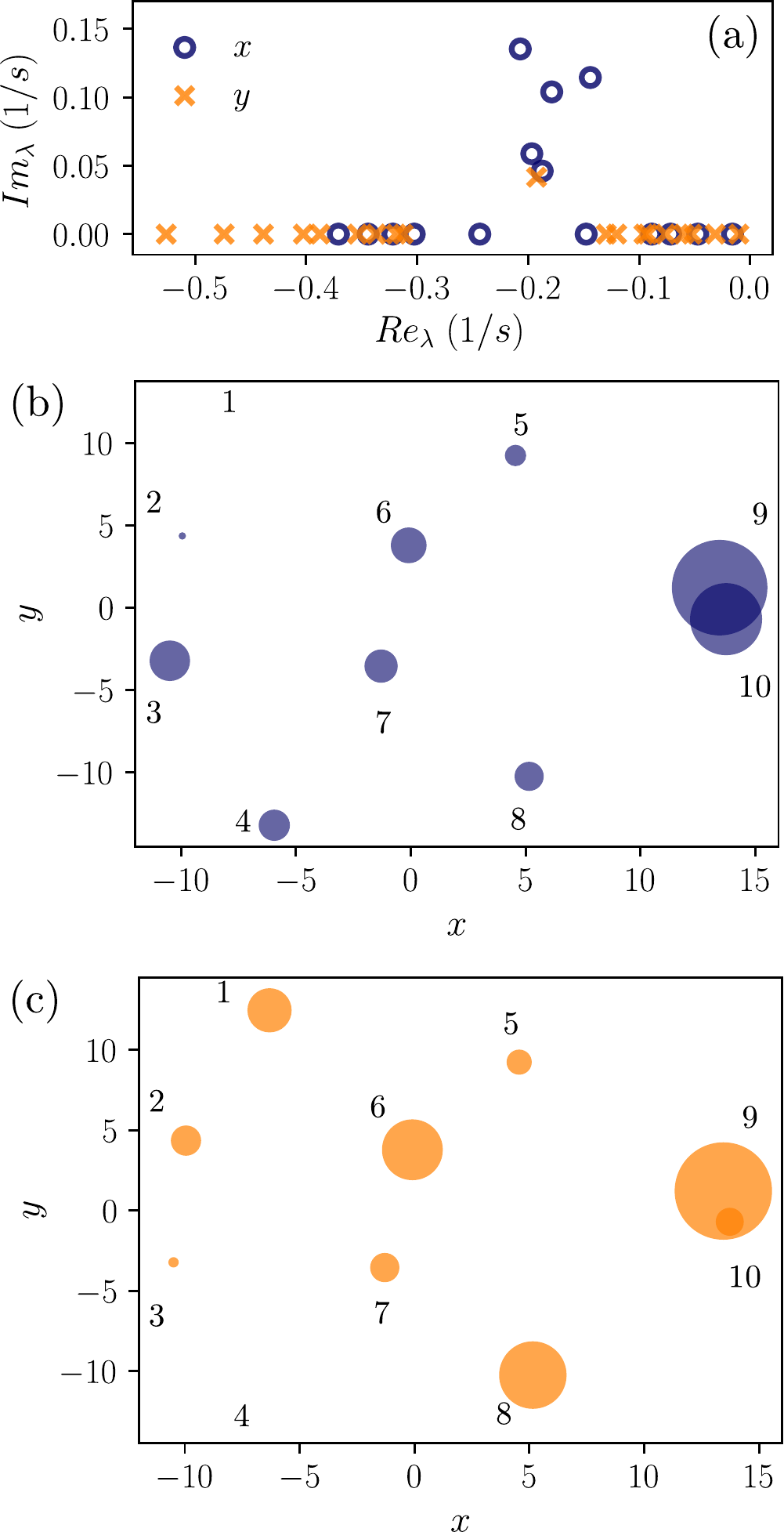}
\caption{
Collective modes. 
(a) Eigenvalues plotted as the real part versus the imaginary part, for the horizontal (circles) and the vertical (crosses) axis.
(b) Visualization of the first mode in the horizontal axis.
(c) Visualization of the first mode in the vertical axis.
See main text for further details.
}
\label{fi:autoval}
\end{figure*}

In Fig.~\ref{fi:autoval} panel $(a)$, we plotted the system's eigenvalues $\lambda \in C$, as $Re_\lambda$ vs. $Im_\lambda$. 
We can see in most cases, $Im(\lambda)=0$. 
However, around $Re(\lambda) \approx -0.2$, we can see the presence of characteristics frequencies in both coordinates.
Let us focus on describe the case of $\lambda_1$, the eigenvalue with smallest real part and $Im(\lambda_1)\ne 0$. 
This case is particularly important, because the energy that enters the system as noise, is transferred mainly to the vibration mode given by the eigenvector associated to $\lambda_1$, $v_{\lambda_1}$ (first mode)
\cite{strogatz2018nonlinear,nayfeh2008applied}.
In the frame of our model, therefore, $v_{\lambda_1}$ bears information on the collective behavior of the players.
We calculated $\lambda^x_1=(-0.14 \pm i~0.11)~(1/s)$ and $\lambda^y_1= (-0.19 \pm i~0.04)~(1/s)$ (notice, complex conjugates are not shown in the plot). 
To describe these collective modes, let us focus on  Fig.~\ref{fi:autoval}.
Here panel $(b)$ is linked to the horizontal coordinate and panel $(c)$ to the vertical one.
In the plots, each circle represents the players in its field's natural positions.
The circles' radii are proportional to the absolute value of the components of $v^x_{\lambda_1}$ (panel $(b)$, blue) and $v^y_{\lambda_1}$ (panel $(c)$, yellow).
Therefore, the size of the circles in the visualization, indicates the effect of the vibration mode on the players, or, in other words, how much the player is involved in this particular collective behavior. 
For instance,  we can see that player $1$ is not affected by the mode in the horizontal coordinate but is highly affected by the mode in the vertical coordinate.
In another example, for the case of player $5$, we can see the collective modes in both coordinates slightly affect the motion of this player. This is because, how we have previously stated, $5$ seems to be a free player in the field, therefore his maneuvers are  not constrained by other players.
Conversely, player $9$ is the most affected in both coordinates; which seems to indicate that the collective behavior of the team directly affects the free movement of this particular player.
A similar analysis can be performed for every team player in the field.

Let us now focus on using this information to describe the behavior of the defenders and their roles in the team.
Fig.~\ref{fi:autoval} shows that defenders 1 and 2, at the left-back in the formation, are slightly involved in the collective mode related to the horizontal axis [panel (B)] and highly involved in the mode related to the vertical axis [panel (C)]. This indicates that these players exhibit a natural trend to coordinate with the team in the vertical direction and behave more freely when they perform movements in the horizontal direction.
Conversely, players 3 and 4, at the right-back in the formation, are highly involved in the collective mode related to the horizontal axis and slightly involved in the mode related to the vertical. This indicates that these defenders exhibit a natural trend to follow the movements of the team in the horizontal axis (towards the goal).
These observations reveal a mixed behavior in the defense, where the defenders 3 and 4 are more likely to participate in attacking actions, whereas defenders 1 and 2 are more likely devoted to covering gaps.
Naturally, an expert coach may easily uncover these kinds of observations while attending a game. 
However, our technique could be useful for the systematic analysis of hundreds or thousands of games.

The reader may note, that the eigenvalues and eigenvectors of the systems provide a handy analytical tool for coaches to assess several aspects of teams dynamics.
As well as in the case of the hierarchical clustering analysis, in this frame, it is possible to link the strengths and weaknesses of a teams to collective modes uncovered by this technique, that could be useful to identify patterns associated to a winning or a losing formation to act accordingly in decision--making processes.

\subsubsection{Using network metrics to analyze the game  Troms{\o}~IL vs. Anzhi}

\begin{figure*}[t!]
\centering
\includegraphics[width=1\textwidth]{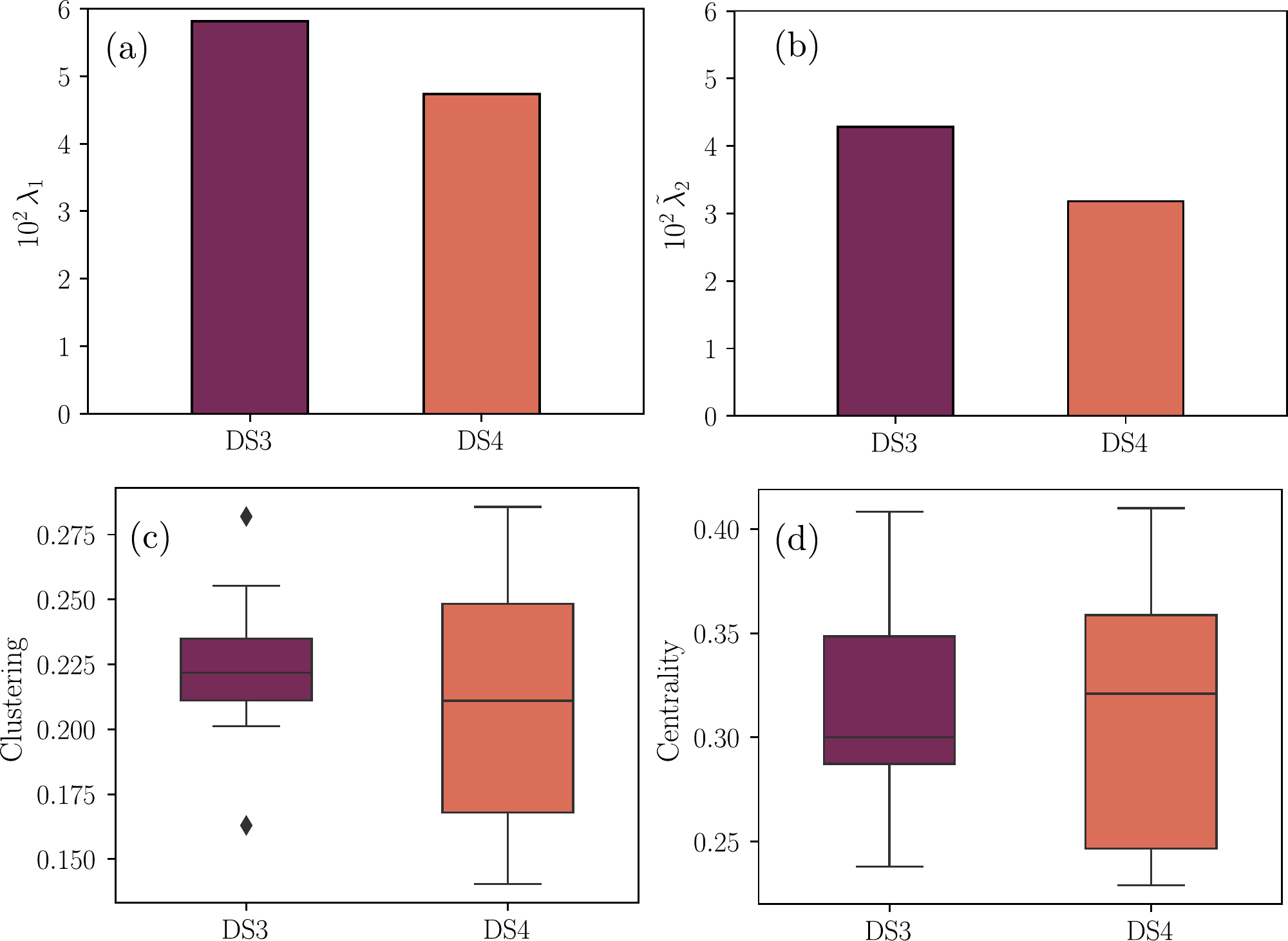}
\caption{
Comparative analysis for the two halves of the game  Troms{\o}~IL vs. Anzhi.
(a) Largest eigenvalue $\lambda_1$ of the Adjacency matrix A.
(b) Algebraic connectivity, $\tilde{\lambda}_2$, of the Laplacian matrix $\tilde{L}$.
(c) Clustering coefficient.
(d) Eigenvector centrality.
}
\label{fi:comparacion}
\end{figure*}

Datasets $DS3$ and $DS4$ are related to the first and the second half of the game Troms{\o}~IL vs. Anzhi. In this game, Anzhi scored a goal at the last minute of the second half to obtain a victory over the local team. 
In this context, the idea is to fit $DS3$ and $DS4$ to the model, obtain the networks of players defined by parameters $\vec{k}_{nm}$ and perform a comparative study of the two cases analyzing our results by using traditional network science metrics.
Let us focus on describing Fig.~\ref{fi:comparacion}.
In panel $(a)$, we show the largest eigenvalue, $\lambda_1$, of the network adjacency matrix in both datasets. 
This parameter gives information on the network strength \cite{aguirre2013successful}. Higher values of $\lambda_1$ indicate that important players in the graph are connected among them.
We can see that $\lambda_1$ in $DS3$ is $\approx 19\,\%$ higher than in $DS4$, which indicates that the network strength decreases in the second half of the game.
In panel $(b)$ we show the algebraic connectivity, $\tilde{\lambda}_2$.
This value corresponds to the smallest eigenvalue of the Laplacian matrix of the players networks \cite{newman2018networks} and bears information on the structural and dynamical properties of the networks. 
Small values of $\tilde{\lambda}_2$ indicate the presence of independent groups inside the network and are also linked to higher diffusion times, thus, it indicates a lack of players' connectivity.
We can see that $\tilde{\lambda}_2$ decreases $\approx 31\,\%$ in $DS4$, which seems to indicate that in the second half of the game the team players lost cohesion.
Box plots presented in panel $(c)$ show the weighted clustering coefficient~\cite{ahnert2007ensemble} of the team players. 
This parameter measures the local robustness of the network. 
We can see that in $DS3$ the clustering is slightly higher than in $DS4$, and the dispersion of the values is lower. 
This indicates that the network of players is more robust and homogeneous in the first half of the game.
Lastly, in panel $(d)$ we show box plots with the eigenvector centrality \cite{newman2008mathematics} of the players in both networks.
The centrality indicates the influence of a player in the team.
A higher value in a particular player is related to strong connections with the other important players in the team.
The mean value of the centrality is in both cases $\approx 0.3$ and the standard deviation $\approx 0.05$. The maximum, likewise, is very similar $\approx 0.4$ and the median, showed in the box plot, is a little lower in $DS3$.
We can also observe that the network linked to $DS3$ seems to exhibits more homogeneous values among the team players than the network linked to $DS4$. 

In the light of the results discussed above, we can see that in the second half of the game the team decreases in connectivity, cohesion, and becomes more heterogeneous. 
Considering that Troms{\o}~IL received a goal at the end of the second half and lost the game, the fall of these particular metrics seems to be related to a decrease in the team's performance.
Notice, previous works devoted to the analysis of passing networks have also reported a relationship between the magnitudes of these particular metrics in these networks and team performances~\cite{buldu2019defining}.
In this regard, the previous analysis suggests that there is consistency between the results reached through our methods and results previously reported in the literature.

\section{Summary and conclusions}

In this work, we studied the spatiotemporal dynamics of a professional football team. 
Based on empirical observations, we proposed to model the player cooperative interactions to describe the global behavior of the group.
In this section, we summarize our main results.

Firstly, we surveyed a database containing body--sensor traces from one team on three professional soccer games.
We observed statistic regularities in the dynamics of the games that reveal the presence of a strong correlation in the players' movements. 
With this insight, we proposed a model for the team's dynamic consisting of a fully connected system where the players interact with each other following linear--spring--like forces.
In this frame, we performed a minimization process to obtain the parameters that fit the model to the datasets.
Furthermore, we showed that is possible to treat the higher--order contributions as stochastic forces in the players' velocities, which we modeled as Gaussian fluctuations. 

Secondly, once defined the model, we carried out numerical simulations and evaluated the model performance by comparing the outcomes with the empirical data.
We showed that the model generates spatiotemporal dynamics that give a good approximation to the real observations. Particularly, we analyzed $(i)$ the probability of finding a player in a position $(x,y)$, $(ii)$ the action zones of the players, and $(iii)$ the temporal structure of the system by studying the time to return to the mean value in the temporal series of the kinetic energy of the system.
Despite its simplicity, in all the cases the model exhibited a good performance. 

Thirdly, we described the system at the local level by using the parameters we obtained from the minimization process.
On this, we proposed to use two analytical tools, a Hierarchical cluster classification, and a Eigenvalues--eigenvectors based analysis.
We found that is possible to describe the team behavior at several organization levels and to uncover non--trivial collective interactions.

Lastly, we used network science metrics to carry out a comparative analysis on the two halves of the game Troms{\o}~IL vs. Anzhi. 
We observed that a decrease in connectivity and cohesion, and an increase in the heterogeneity of the network of players, seem to be related to a decrease in the team performance.

We consider this contribution is a new step towards a better understanding of the game of football as a complex system. 
The proposed stochastic model, based on linear interactions is simple and can be easily understood in the frame of standard dynamical variables.
Moreover, our framework provides a handy analytical tool to analyze and evaluate tactical aspects of teams, something helpful to support the decision-making processes that coaches face in their working activities.
It is important to highlight that our framework is not limited to be used only in the analysis of football games. 
A similar approach can be performed to study
others sport disciplines, mainly when the evaluation of players' interactions is key to understand the game results.
In \cite{cervone2016multiresolution}, for instance, the authors use players-tracking data in basketball games to estimate the expected number of points obtained by the end of a possession. 
In this case, a complementary analysis within our framework could also unveil collective behavior patterns linked to players' coordination interactions that can be correlated to the upshot at the end of the possession intervals.

However, we point out that the full dynamics of a football match (and others sports) cannot be addressed by only analyzing cooperative aspects within a particular team. 
To describe the full dynamics, we should also model the competitive interactions among the players in both teams.
To do so, we need to measure the interplay among rivals, which in our framework implies having body--sensor traces for both teams, something that football clubs reject because of competing interests.
Moreover, it could be useful to have a record of the ball position to improve our analysis.
In this context, it becomes relevant the use of alternatives measurement techniques base on artificial intelligence and visual recognition \cite{sanford2020group,ganesh2019novel,khan2018soccer}.

To summarize, our model provides a simple approach to describe the collective dynamics of a football team untangling interactions among players, and stochastic inputs. The structure of interactions that turns
out from this approach can be considered a new metric for this sport.
In this sense, our analysis complement recent contributions in the framework of network science~\cite{martinez2020spatial,herrera2020pitch,buldu2019defining,gonccalves2017exploring,garrido2020consistency,narizuka2021networks}.
Note, there is a major difference between our approach and current network-science-oriented methods: the latter analyzes interaction based on players' passes, whereas our approach analyzes interaction based on players' movements.
In this regard, the problem of study a football team from only passing/pitch networks is that this approach is entirely based on on-ball action, which completely neglects how players behave when they are far from the center of the plays (off-ball actions). 
Our approach, instead, integrates the information of the entire team to calculate every single link between players, therefore in our model, we are also considered off-ball actions, which is key to correctly evaluate teams' performance~\cite{casal2017possession}.
In addition, many of the metrics of current use in this sport can be tested through our model~\cite{clemente2013collective,frencken2011oscillations,moura2013spectral,yue2008mathematicalp1,yue2008mathematicalp2,narizuka2019clustering}.
On the other hand, to perform an analysis based on passing/pitch networks, it is required to have a record of the ball position and be able to characterize events (passes) during the match. 
Our model, instead, employs just the data of players' positions, something that nowadays can be easily measured, both in training sessions and competitive scenarios, with the simplest GPS--trackers available in the market. 

Finally, we consider that is possible to improve our model by incorporating non--linear interactions in the equations of motion.
An analysis of the most commonly observed players' maneuvers may help to find new statistical patterns to be used as an insight to propose a non--linear approach. In this regard, we leave the door open to futures researches projects in the field.

\section*{Acknowledgement}
This work was partially supported by grants from CONICET (PIP 112 20150 10028), FonCyT (PICT-2017-0973), SeCyT–UNC (Argentina) and MinCyT Córdoba (PID PGC 2018).

\clearpage
\newpage

\begin{table}[h!] 
\centering
\begin{tabular}{ || c | c | c || c | c | c || c | c | c || c | c | c || } 
\hline
Par. & x & y & Par. & x & y & Par. & x & y & Par. & x & y \\ 
\hline 
$\alpha_1$ & $4.0\cdot10^{-1}$ & $4.3\cdot10^{-1}$ & $k_{12}$ & $4.2\cdot10^{-3}$ & $1.2\cdot10^{-2}$ & $k_{37}$ & 0.0 & $2.5\cdot10^{-3}$ & $k_{79}$ & 0.0 & $5.1\cdot10^{-4}$  \\ 
$\alpha_2$ & $4.1\cdot10^{-1}$ & $5.8\cdot10^{-1}$ & $k_{13}$ & 0.0 & 0.0 & $k_{38}$ & 0.0 & 0.0 & $k_{7\,10}$ & $4.5\cdot10^{-3}$ & $1.4\cdot10^{-3}$  \\ 
$\alpha_3$ & $3.8\cdot10^{-1}$ & $5.7\cdot10^{-1}$ & $k_{14}$ & $6.4\cdot10^{-3}$ & 0.0 & $k_{39}$ & $6.4\cdot10^{-3}$ & 0.0 & $k_{89}$ & $5.6\cdot10^{-3}$ & $6.9\cdot10^{-3}$  \\ 
$\alpha_4$ & $4.0\cdot10^{-1}$ & $4.2\cdot10^{-1}$ & $k_{15}$ & $3.9\cdot10^{-3}$ & $4.7\cdot10^{-3}$ & $k_{3\,10}$ & 0.0 & $2.3\cdot10^{-3}$ & $k_{8\,10}$ & $6.0\cdot10^{-3}$ & $3.2\cdot10^{-3}$  \\ 
$\alpha_5$ & $4.1\cdot10^{-1}$ & $4.1\cdot10^{-1}$ & $k_{16}$ & $1.7\cdot10^{-3}$ & $6.7\cdot10^{-3}$ & $k_{45}$ & $1.7\cdot10^{-3}$ & 0.0 & $k_{9\,10}$ & $4.7\cdot10^{-3}$ & 0.0  \\ 
$\alpha_6$ & $4.1\cdot10^{-1}$ & $5.4\cdot10^{-1}$ & $k_{17}$ & $4.6\cdot10^{-3}$ & 0.0 & $k_{46}$ & 0.0 & $1.8\cdot10^{-3}$ &  - & - & -     \\ 
$\alpha_7$ & $4.2\cdot10^{-1}$ & $5.0\cdot10^{-1}$ & $k_{18}$ & $1.9\cdot10^{-3}$ & $2.1\cdot10^{-3}$ & $k_{47}$ & 0.0 & $5.4\cdot10^{-3}$ &  - & - & -     \\ 
$\alpha_8$ & $4.0\cdot10^{-1}$ & $4.4\cdot10^{-1}$ & $k_{19}$ & $3.7\cdot10^{-3}$ & $1.4\cdot10^{-3}$ & $k_{48}$ & $1.8\cdot10^{-3}$ & $3.3\cdot10^{-3}$ &  - & - & -     \\ 
$\alpha_9$ & $2.8\cdot10^{-1}$ & $3.3\cdot10^{-1}$ & $k_{1\,10}$ & $5.0\cdot10^{-3}$ & $1.6\cdot10^{-3}$ & $k_{49}$ & $3.6\cdot10^{-3}$ & $6.5\cdot10^{-4}$ &  - & - & -     \\ 
$\alpha_{10}$ & $2.8\cdot10^{-1}$ & $3.7\cdot10^{-1}$ & $k_{23}$ & $6.7\cdot10^{-3}$ & $7.1\cdot10^{-3}$ & $k_{4\,10}$ & 0.0 & $1.1\cdot10^{-3}$ &  - & - & -     \\ 
$k_{a1}$ & $5.7\cdot10^{-3}$ & 0.0 & $k_{24}$ & 0.0 & $8.9\cdot10^{-4}$ & $k_{56}$ & $2.9\cdot10^{-3}$ & $6.9\cdot10^{-4}$ &  - & - & -     \\ 
$k_{a2}$ & 0.0 & 0.0 & $k_{25}$ & $2.8\cdot10^{-6}$ & 0.0 & $k_{57}$ & 0.0 & 0.0 &  - & - & -     \\ 
$k_{a3}$ & 0.0 & $2.2\cdot10^{-2}$ & $k_{26}$ & $2.1\cdot10^{-3}$ & $6.7\cdot10^{-3}$ & $k_{58}$ & $8.0\cdot10^{-3}$ & 0.0 &  - & - & -     \\ 
$k_{a4}$ & $1.0\cdot10^{-2}$ & $1.0\cdot10^{-2}$ & $k_{27}$ & $5.5\cdot10^{-3}$ & 0.0 & $k_{59}$ & $1.6\cdot10^{-3}$ & $2.7\cdot10^{-3}$ &  - & - & -     \\ 
$k_{a5}$ & $2.0\cdot10^{-2}$ & 0.0 & $k_{28}$ & 0.0 & $9.9\cdot10^{-4}$ & $k_{5\,10}$ & 0.0 & $2.1\cdot10^{-3}$ &  - & - & -     \\ 
$k_{a6}$ & $2.7\cdot10^{-2}$ & $2.2\cdot10^{-2}$ & $k_{29}$ & $1.9\cdot10^{-3}$ & $5.4\cdot10^{-3}$ & $k_{67}$ & $8.6\cdot10^{-3}$ & $5.0\cdot10^{-3}$ &  - & - & -     \\ 
$k_{a7}$ & $3.2\cdot10^{-2}$ & $1.2\cdot10^{-2}$ & $k_{2\,10}$ & $1.5\cdot10^{-3}$ & $3.5\cdot10^{-3}$ & $k_{68}$ & $2.2\cdot10^{-4}$ & $9.9\cdot10^{-5}$ &  - & - & -     \\ 
$k_{a8}$ & $1.9\cdot10^{-2}$ & $1.2\cdot10^{-2}$ & $k_{34}$ & 0.0 & $3.3\cdot10^{-3}$ & $k_{69}$ & 0.0 & $5.4\cdot10^{-3}$ &  - & - & -     \\ 
$k_{a9}$ & $4.2\cdot10^{-4}$ & 0.0 & $k_{35}$ & 0.0 & 0.0 & $k_{6\,10}$ & $3.2\cdot10^{-3}$ & $1.9\cdot10^{-3}$ &  - & - & -     \\ 
$k_{a10}$ & 0.0 & 0.0 & $k_{36}$ & $1.6\cdot10^{-3}$ & $4.0\cdot10^{-3}$ & $k_{78}$ & 0.0 & $5.7\cdot10^{-3}$ &  - & - & -     \\ 
\hline 
\end{tabular} 
\caption{
Model parameters inferred for dataset $DS1$.
} 
\label{tabla1} 
\end{table} 

\begin{figure*}[t!]
\centering
\includegraphics[width=\textwidth]{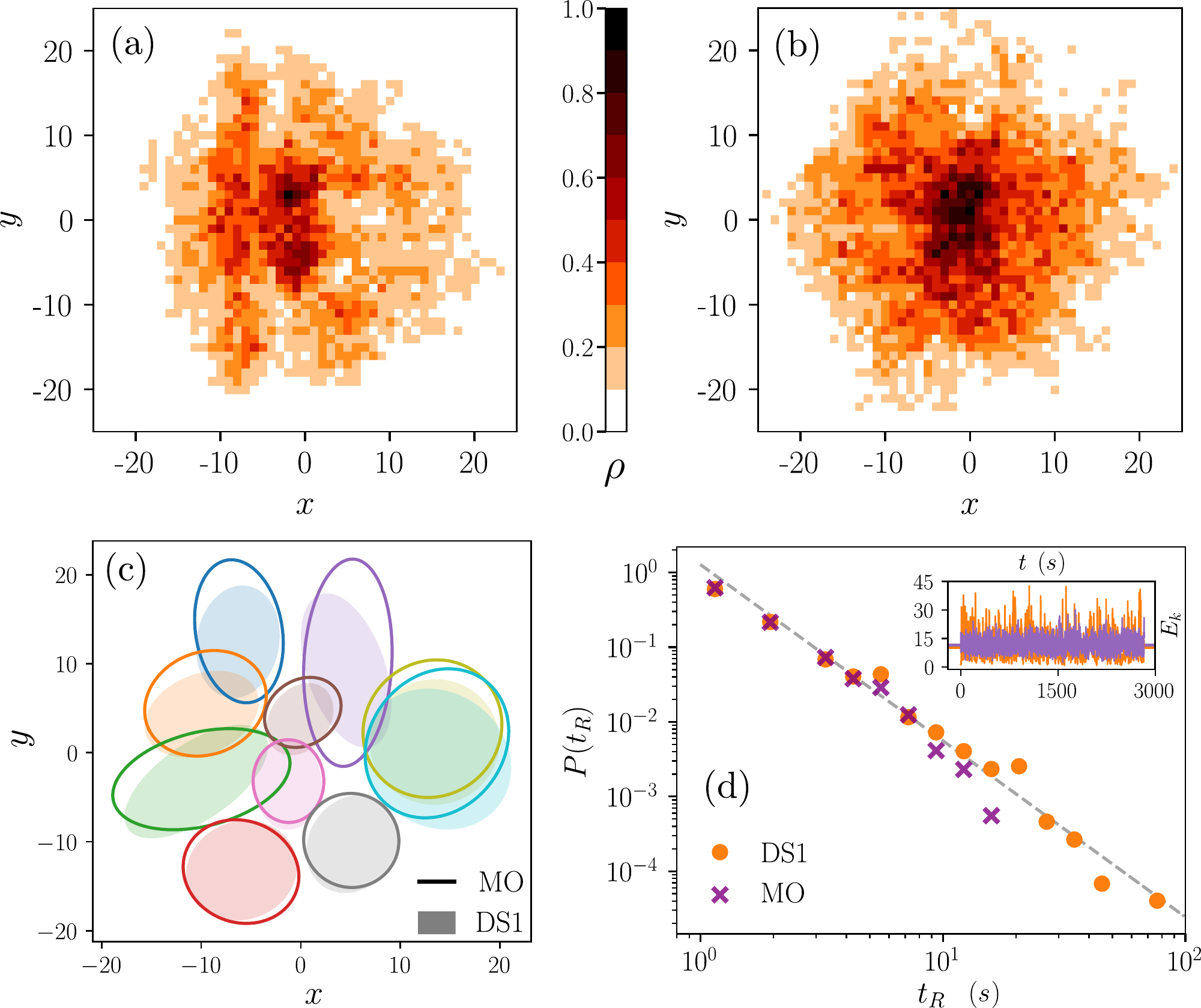}
\caption{
Results on the collective dynamics simulations.
Panels (a) and (b), probability of finding a player in the field, in empirical data and simulations, respectively. 
(c) Players' action zones. Empirical data (shadow areas) compared with simulations (curves).
(d) Probability distribution of the time to return to the mean value, $P(t_R)$. The dashed line indicates a non--linear 
fit performed to the empirical data (orange circles). The inset shows the evolution of the kinetic energy, $E_k(t)$ from where $t_R$ is measured.
}
\label{fi:simulations}
\end{figure*}

\end{document}